# Can we constrain the origin of Mars' recurring slope lineae using atmospheric observations?

Hiroyuki Kurokawa[1,*,†], Takeshi Kuroda[2,†], Shohei Aoki[3,4], Hiromu Nakagawa[2]

1: Earth-Life Science Institute, Tokyo Institute of Technology, Tokyo, Japan
2: Department of Geophysics, Tohoku University, Sendai, Japan
3: Institute of Space and Astronautical Science, Japan Aerospace Exploration Agency, Kanagawa, Japan
4: Royal Belgian Institute for Space Aeronomy, Brussels, Belgium

*Corresponding author. hiro.kurokawa@elsi.jp
†Hiroyuki Kurokawa and Takeshi Kuroda equally contributed to this work.

## Abstract

Flowing water and brine have been proposed to cause seasonally reappearing dark streaks called recurring slope lineae (RSL) on steep warm slopes on Mars, along with other formation mechanisms that do not involve water. This study aims to examine whether the evaporation of water vapor from the RSL, whether from fresh water or brine, is detectable by observing water vapor and/or clouds. In this study, we summarize the possible rate and duration of water-vapor emission from RSL in different scenarios, simulate how the emitted water vapor behaves in a global climate model, and discuss the detectability of water vapor in nadir observations during existing and future explorations. We found that, in typical cases, rapid horizontal dissipation within the planetary boundary layer (PBL) following the release of water vapor prohibits cloud formation and the excess water vapor from being distinguished from the background with existing observations. Thus, we conclude that the lack of correlation between the RSL activities and the overlying water-vapor column density does not necessarily rule out the wet origin of RSL. Nevertheless, we also found that water vapor tends to accumulate in basins and valleys in some cases due to the combined effects of topography and low PBL; we suggest the locations of such configuration as targets for future atmospheric studies of Mars dedicated to quantifying water-vapor release (associated with RSL) to elucidate the formation mechanism(s) of the RSL on the planet.

**Keywords**: Mars, surface; Mars, atmosphere; Mars, climate; Atmospheres, dynamics

## 1. Introduction

Water on the surface of present-day Mars has attracted great attention for a long time as a record of the history of the surface environment, target for astrobiological investigations, and a possible resource for future human exploration missions. High-resolution images taken by the High-Resolution Imaging Science Experiment (HiRISE) on the Mars Reconnaissance Orbiter (MRO) found a new type of dark streak on steep slopes called "recurring slope lineae" (RSL) (McEwen et al. 2011). RSL are lengthening streaks on Mars that reappear each year during the warm seasons and fade during the cold seasons; they are possibly linked to temperature and/or solar irradiation on the surface of the planet.

Liquid water and brine have been proposed as the flowing material on RSL itself (McEwen et al. 2011, 2014; Grimm et al. 2014; Stillman et al. 2014, 2016, 2017) or as a trigger of the granular flow (e.g., melting and boiling) (Massé et al. 2016; Raack et al. 2017). Various sources of water have been proposed, including subsurface aquifers, melting of ice dams, and deliquescent salt (Ojha et al. 2015; Wang et al. 2019). In contrast, several studies have proposed that RSL are dry granular flows triggered by liquid-free mechanisms, such as solar irradiation-induced Knudsen pump, and that the albedo change is caused by the sorting of grains and removal/deposition of dust (Dundas et al. 2017; Dundas 2020; Schmidt et al. 2017; Schaefer et al. 2019).

Previous studies that have tried to constrain the formation mechanisms of RSL have mainly focused on surface features by using visible images, infrared spectra, and thermal data, to observe surface features (see the following three paragraphs for references). The wet-origin hypothesis is based mainly on the correlation between the RSL activity and local temperatures. McEwen et al. (2011, 2014) first analyzed the HiRISE data to map the RSL locations and time and reported extensive RSL activity in the warm seasons in the equatorial regions of Mars, suggesting a connection to the seasonal melting of frozen brines. Stillman et al. (2014, 2016, 2017) combined the HiRISE images of different RSL areas, with the surface temperatures derived from the Mars Global Surveyor (MGS) Thermal Emission Spectrometer (TES), MRO Mars Climate Sounder (MCS), and Mars Odyssey (MO) Thermal Emission Imaging System (THEMIS) data. They found that the southern mid-latitude (SML) RSL are consistent with freshwater subsurface flows, whereas those in Valles Marineris (VM) and northern Chryse Planitia and southwestern Acidalia Planitia (CAP) are consistent with brines. We note that an updated study found that the SML RSL starts lengthening earlier than what was previously thought (Stillman and Grimm 2018). Abotalib and Heggy (2019) found a correlation between the source regions of the RSL and multi-scale fractures and argued a deep groundwater origin, although this may not be applicable to all RSL because RSL are also found among broad elevation ranges, including the tops of isolated peaks and ridges (Chojnacki et al. 2016). Observations of anomalous RSL (early faders, collinear RSL, and those that emerge at featureless locations) associated closely with standard RSL are more consistent with subsurface liquid flow in porous regolith than with dry granular flow (Lark et al. 2021).

Additionally, the presence of hydrated salts, if identified, may be an indicator of a wet origin of RSL. Ojha et al. (2015) reported the detection of hydrated salts using a Compact Reconnaissance Imaging Spectrometer for Mars (CRISM) onboard the MRO for the four locations that they studied. However, subsequent studies found the signal likely due to an artifact in the CRISM data (Leask et al. 2018; Vincendon et al. 2019).

Other studies have reported supporting evidence for the dry (liquid-free) origin of RSL. RSL on unconsolidated Martian sand are found primarily on slopes at the angle of repose, indicating a dry mechanism (Dundas et al. 2017); however, more recent studies have shown that some RSL reach slopes below the angle of repose (Tebolt et al. 2020; Stillman et al. 2020). RSL formation and growth were enhanced during the aftermath of a global dust storm (McEwen et al. 2021). Vincendon et al. (2019) showed that the location and timing of flow activity are consistent with the removal and deposition of brighter dust on darker surfaces in relation to seasonal dust storm activity. Edwards and Piqueux (2016) argued that the lack of temperature difference in the THEMIS data between RSL-bearing and RSL-free terrains in the Garni crater in the VM limits the water and salt contents. Inconsistency between the temperature and RSL activity has also been reported for a smaller scale at several RSL sites by combining local thermal simulations (Millot et al. 2021). Furthermore, by combining the images taken by the Color and Surface Science Imaging System (CaSSIS) on board the ESA (European Space Agency) ExoMars Trace Gas Orbiter (TGO) with HiRISE, Munaretto et al. (2020) found limited albedo changes at different local hours for the RSL at the Hale crater and concluded that these RSL were not caused by seeping water that reached the surface.

Overall, the formation mechanism(s) of RSL remains unsolved and may vary depending on the RSL site. Thus, further observational constraints are important for understanding the formation mechanism of RSL and their possible relationship with water.

In contrast to previous studies, we focus on the possibility of constraining the origin of RSL using atmospheric observations. Even if RSL have a wet origin, liquid water or brine may not be present on the surface if it is just a trigger of granular flow or if propagation of liquid occurs mainly in the subsurface by intergranular percolation (Grimm et al. 2014; Massé et al. 2016). Notably, even in these cases, the evaporation of water releases water vapor to the overlying atmosphere. The detection of water vapor evaporated from the RSL sites may thus, provide independent constraints on the formation mechanism. Moreover, water ice clouds and fog are potential signatures of water released from RSL sites. Leung et al. (2016) proposed that the water vapor released from the RSL at VM may aid the formation of fog that may have been detected within VM (Möhlmann et al., 2009), although spectral analysis showed that the observed structures are most likely dust hazes rather than water ice fog (Inada et al. 2008).

This study examines whether the formation mechanism of RSL can be constrained by observing atmospheric water vapor or water ice clouds. In Section 2, we summarize the expected rate and duration of water-vapor emissions for the proposed mechanisms of RSL formation. Section 3 presents the simulations using a Mars Global Climate Model (MGCM) to explain how the emitted water vapor is transported in the atmosphere and whether water vapor saturates to form water ice clouds. Section 4 discusses the implications for the origin of the RSL and existing/future atmospheric observations. Finally, Section 5 concludes the paper.

## 2. Formation mechanisms of recurring slope lineae (RSL) and water-vapor emission

We first summarize what is known for water vapor in the background (without possible release from RSL) atmosphere (Section 2.1) and then discuss possible emissions from RSL sites (Section 2.2).

## 2.1 Water vapor in the background atmosphere

It is well known that a small amount of water vapor is always present in the Martian atmosphere (~10 pr.um, Smith 2008). However, its abundance varies across time and space. The global seasonal variation of water-vapor column integrated abundance is controlled by sublimation/condensation from/to the polar ice caps and the formation of water ice clouds due to the large seasonal changes in insolation over the globe and the resultant general circulation (Montmessin et al. 2004). Water column density peaks near 60–70 pr.µm in the northern summer at the polar region, whereas only 20 pr.µm is observed during the southern summer at the southern pole. The equivalent volume of ice sublimating as vapor in the atmosphere is up to 2.3 km$^3$ at Ls = 100° (at the maximum water-vapor abundance) on a global average (Montmessin et al., 2017). After sublimation from the cap, water vapor is transported toward the equator in the summer season; therefore, this season is recognized as the origin of the equatorial cloud belt. During the fall and winter of both hemispheres, the concentrations of water vapor at high latitudes decrease dramatically (Smith 2008). The observations also revealed the planetary scale geographical distribution at the low-middle latitudes: two local maxima located over Tharsis and Arabia Terra due to atmospheric dynamics or possible release of water from the subsurface (Fouchet et al. 2007). In addition, diurnal variations in the water-vapor mixing ratios at the surface (from 10–30 ppm to 50–80 ppm) imply adsorption of water with the Martian regolith, as per the measurements obtained by the Curiosity rover at Gale crater (Savijärvi et al. 2016). However, the global role of water exchange between the planet's regolith and atmosphere is still not understood.

## 2.2 Possible water-vapor release from RSL

RSL are narrow (0.5–5 m) but long (> 50 m), and occur as a group of ~100 in a single RSL site (McEwen et al. 2011, 2014; Stillman et al. 2014). The RSL sites are mainly located in four primary RSL regions, though they are not limited to them (Bhardwaj et al. 2019): SML, VM, CAP, and equatorial highlands (EH) (Stillman et al. 2014). The surficial coverage of the RSL at an RSL site is 10 % or less.

The seasonal variation in RSL can be typically classified into four periods: (i) lengthening, (ii) stationary, (iii) fading, and (iv) inactive (Figure 1; Stillman et al. 2014). However, in rare cases, RSL has also been observed to fade and grow simultaneously on the same slope (Stillman et al. 2017, 2020; Leak et al. 2021). The first period and last two periods correspond to the warm and cold seasons, respectively. The critical temperature required to initiate lengthening is dependent on the RSL region (Stillman et al. 2014, 2016, 2017). SML-RSL are found to start lengthening at 273 K and are suggested to be formed by shallow subsurface, pure liquid water (Stillman et al. 2014). CAP- and VM-RSL start to develop at 238 K and 246–264 K, respectively, and are considered to be associated with brine (Stillman et al. 2016, 2017). We note that the correlation of RSL activity with surface temperature did not apply to all RSL (Vincendon et al. 2019; Stillman et al. 2020).

To estimate the rate and duration of water-vapor evaporation from the RSL, we classified the proposed formation mechanisms as follows:

(a) Wet origin: RSL were originally thought to be formed by the flow of water or brine on the surface or in the shallow subsurface (McEwen et al. 2011, 2014; Grimm et al. 2014; Stillman et al. 2014, 2016, 2017; Abotalib and Heggy 2019; Imamura et al. 2019; Huber et al. 2020; Leak et al. 2021). In this scenario, RSL are wet, and thus, water evaporates through the periods when they are dark and thus visible, namely in periods (i) lengthening to (iii) fading. Their low albedo can be caused by water (Zhang and Voss 2006; Massé et al. 2014). We note that, although we focus on evaporation, water vapor can also be taken up from the atmosphere, owing to condensation, regolith adsorption (Zent et al. 1993; Steele et al. 2017; Savijärvi et al. 2016; 2019; Hu 2019), or deliquescence of salts (Martín-Torres et al. 2015).

(b) Hybrid (dry but wet-triggered) origin: Several studies have proposed that RSL are granular flows triggered by mechanisms associated with water, such as boiling water (Massé et al. 2016; Raack et al. 2017), deliquescence of subsurface salts (Wang et al. 2019; Bishop et al. 2021), and changes in soil cohesion due to the loss of water (Shoji et al. 2019; Gough et al. 2020). In Cases (b) and (c), the albedo change is not associated with water or brine, but the debris flow typically removes light-colored surface grains and exposes darker underlying grains, causing the darkening of RSL (Dundas et al. 2017; Dundas 2020; Schmidt et al. 2017; Schaefer et al. 2019; Vincendon et al. 2019). Water vapor is released into the atmosphere during period (i) lengthening; however, emissions are not necessarily expected for periods (ii)–(iv).

(c) Dry origin: RSL are also proposed to form completely liquid-free mechanisms, and they are thus, associated with neither water nor brine. In this scenario, solar irradiation-induced Knudsen pump and dust loading by dust storms cause RSL formation (Dundas et al. 2017; Dundas 2020; Schmidt et al. 2017; Schaefer et al. 2019; Vincendon et al. 2019). As with Case (b), the albedo change is associated with the removal/deposition of dust.

The emission continues for periods (i)–(iii) in Case (a) and for period (i) in Case (b); there is no emission in Case (c). Because the surface of the RSL is wet in Case (a), the minimum estimate for the emission rate of water vapor is given by the evaporation rate from the total surface area of the RSL. The evaporation rate of pure water highly depends on the temperature and is ~1 mm h$^{-1}$ at 273 K (Murphy and Koop 2005), whereas that of brine is a weak function of temperature (Altheide et al. 2009; Chevrier et al. 2009; Hanley et al. 2012). A comparison of temperatures for the onset of CAP-RSL activity and for the eutectic points of salts indicates chloride (Cl$^-$), chlorate (ClO$_3^-$), and/or perchlorate (ClO$_4^-$) brines (Stillman et al. 2016). Experiments have shown that evaporation rates of these brines range 0.1–0.5 mm h$^{-1}$ (depending on the salinity and temperature, Altheide et al. 2009; Chevrier et al. 2009; Hanley et al. 2012). This range of evaporation rate is consistent with the values that reproduce the observed properties of RSL in a numerical model for Case (a) (Grimm et al. 2014). In Case (b), the total amount of water is likely to be smaller than that in Case (a), although the actual value depends on the trigger mechanisms and is difficult to estimate. For instance, the "levitation" induced by boiling water (Raack et al. 2017) reduces the required water volume by a factor of 10.

To summarize, the water-vapor emission rate expected from wet-origin RSL is ~0.1–1 mm h$^{-1}$ (Section 2.2). If water vapor piles up locally, the resulting precipitable column height easily exceeds the background value (a few tens of pr.µm; Section 2.1). Therefore, deviation from the global circulation pattern in local regions would indicate the sources of water vapor near the surface. In Section 3, we study how water vapor released from the RSL is transported in the atmosphere, and whether water ice clouds can form from the emitted water vapor.

## 3.    Global climate model (GCM) simulations and analyses

### 3.1 Settings of the simulations

Given the estimate of water-vapor emission rate and duration, we carried out MGCM simulations to assess how the emitted water would be transported in the atmosphere to determine whether it would be visible in the water-vapor observations. We used a spectral MGCM named DRAMATIC (Dynamics, RAdiation, MAterial Transport, and their mutual InteraCtions) having a horizontal resolution of ~1.1° (or ~67 km) for longitude and latitude (Kuroda et al., 2015, 2016, 2019, 2020); the water vapor emitted from the RSL was added to the lowest layer of the grid points where RSL exist in the MGCM. The released water vapor is then, transported by convection, owing to atmospheric instability and horizontal wind advection. We note that condensation of the emitted water vapor was not considered, but we confirmed that water vapor never reaches saturation in our simulation setup (Section 3.3). The amount of emitted water vapor was averaged in a model grid (size of ~67 km × ~67 km on the equator and proportional to the cosine of latitude), and the subgrid variances smaller than this scale could not be expressed in this MGCM. Nevertheless, even with the horizontal grid size being larger than that of an RSL site, we could investigate how the water emitted from the RSL was transported over hundreds of kilometers, which helps us to understand the physical processes affecting the observational detectability of emitted water. The details of the model and the implemented physical processes are described by Kuroda et al. (2005, 2013). The vertical coordinate is 53 $\sigma$-levels, with a top altitude of ~80-100 km as shown in Table 1 of Kuroda et al. (2016) with increased lower layers below $\sigma = 0.5$, with $\sigma = 0.999$ at the top of the lowest layer (i.e., the thickness of the lowest layer is ~10 m, and the scale height of the Martian atmosphere is ~10 km).

We implemented the water-vapor emission from the RSL on this MGCM to investigate the observability of the emitted water. Figure 2 shows the five emission points (CAP, VM, SML1, SML2, and SML3) in the MGCM chosen from the observed representative RSL (Stillman et al. 2014, 2016, 2017).

Six independent simulations were performed using the water emission from one of the emission points. The onset solar longitudes ($L_s$) of the emissions were set at $L_s = 50°$ for the CAP point, $L_s = 130°$ and $320°$ for the VM point, and $L_s = 289°$ for the three points of the SML (SML1, SML2, and SML3), according to the seasons. The onsets of RSL were observed at each location. The emission rates of water vapor from RSL (hereafter referred to as $F_w$) were given as 0.1 mm h$^{-1}$ at the CAP/VM points (indicating salty water) and 1 mm h$^{-1}$ at the SML points (indicating pure water) (Murphy and Koop 2005; Altheide et al., 2009; Chevrier et al. 2009; Hanley et al. 2012). The evaporation rate of brine can vary by a factor of a few depending on the salt type (Altheide et al., 2009; Chevrier et al. 2009; Hanley et al. 2012). We discuss the influence of the choice of evaporation rates in Section 4. The emissions from the RSL were set to occur when the simulated surface temperature $T$ was higher than the threshold surface temperature (hereafter $T^*$; 238 K at the CAP/VM points and 273 K at the SML points). This setup corresponds to (a) the wet-origin scenario (Section 2.2), in which the duration of water-vapor emission is the longest among the three scenarios; however, we discuss the implications for (b) the hybrid origin scenario in Section 4. The RSL area (source of water emission) at each point was assumed to be 0.1 km$^2$, which is a typical size of a RSL area as seen in an image taken by HiRISE (Figure 3 of McEwen et al. 2014). The dust opacity and its seasonal change in the simulations were defined using the "MY26 dust scenario" (MY represents "Mars Year") (Montabone et al., 2015).

During the execution of these simulations, we need to check the validity of surface temperatures, especially in daytime, simulated in the MGCM. Figure 3 shows the simulated daytime (~2 pm) surface temperatures of the MGCM in comparison with the MGS-TES observations in MY26. The simulated surface temperatures were generally higher than the MGS-TES observations (up to ~20 K). The exceptions are the spring and summer poles. This is because of slower recessions of seasonal polar ice caps in the MGCM. Nevertheless, at the chosen pairs of season and latitude for the RSL simulations, the discrepancies were less than ~8 K. Because we fixed the evaporation rate of water vapor during the emission, the temperature discrepancies do not directly influence water vapor emission in our simulations.

## 3.2 Results of the simulations

Simulations with water-vapor emissions were performed for 20 sols for each point and period. First, we show the theoretical estimations of the quantity of water-vapor emissions that can be reproduced in the MGCM. The grid size ($A$) of the emitted points is given by the following formula:

$$A \approx \frac{2\pi^2 a^2 cos\phi}{N_{\text{lon}} N_{\text{lat}}},$$ (1)

where $a$ ($a = 3396$ km) is the radius of Mars, $N_{\text{lon}}$ ($N_{\text{lon}} = 320$) is the number of longitudinal grid points in the MGCM, $N_{\text{lat}}$ ($N_{\text{lat}} = 160$) is the number of latitudinal grid points, and $\varphi$ is the latitude. This estimation is not strict because the latitudinal grids are defined by the Gaussian grid (Washington and Parkinson 2005), whose intervals are not constant in the spectral global model.

The amount of injected water vapor converted into column density at the grid point of the water source, $\Delta\alpha$, and volume mixing ratio at the lowest layer of the grid point, $\Delta q_{v1}$, can be estimated as follows:

$$\Delta\alpha \approx F_{\text{w}} \cdot \frac{0.1 \text{ km}^2}{A} \cdot \Delta t,$$ (2)

$$\Delta q_{v1} \approx \Delta\alpha \cdot \frac{g\rho_{\text{w}}}{\Delta p_1} \cdot \frac{M_{\text{CO2}}}{M_{\text{H2O}}},$$ (3)

where $g$ ($g = 3.71$ m s$^{-2}$) is the gravitational acceleration, $\rho_{\text{w}}$ ($\rho_{\text{w}} = 1000$ kg m$^{-3}$) is the density of liquid water, $\Delta p_1$ ($\Delta p_1 = 0.001$ $p_{\text{s}}$) is the thickness of the lowest layer in our pressure grid, $p_s$ is the surface pressure, $M_{CO2} = 44$ amu and $M_{H2O} = 18$ amu are the molecular weights of $CO_2$ and $H_2O$, respectively, and $\Delta t$ is the duration of the emission (or $T > T^*$). Table 1 shows the estimated $\Delta \alpha$ and $\Delta q_{v1}$ for each emission point. The values of $\Delta \alpha$ and $\Delta q_{v1}$ are the emitted amount of water from sources per sol, without considering the spreading by vertical and horizontal transport. Although $\Delta \alpha$ listed here were smaller than the "background," the water-vapor column density observed by MGS-TES (Section 2.1; Smith 2008) by 2–3 orders of magnitude due to the horizontal grid size being larger than that of a RSL site, the actual values at the emission spots can be much higher when investigated on a smaller scale (e.g., an order of kilometers). The purpose of our simulations is to investigate the behavior of released water vapor, for example, whether water vapor dissipate or accumulate. $\Delta \alpha$ and $\Delta q_{v1}$ would become comparable to those of background water vapor if water vapor would have accumulated for several tens of sols. We note that we consider the loss of water vapor to the surface neither by regolith adsorption nor by deliquescence of salts, and thus water vapor emitted in our MGCM simulations should be considered as an upper limit.

Figure 4 shows the maximum grid-mean column densities of emitted water vapor in each run of our simulations (hereafter referred to as $\alpha_{max}$), and the ratios of $\alpha_{max}$ against the background water-vapor column densities taken from the MGS-TES observations in MY26 (Smith, 2008) (hereafter referred to as $r_{max}$) at the corresponding season and place. The grid points, which have maximum values in each run, were mostly the same as those of water sources during daytime with emissions, but may vary during nighttime (when there are no emissions). The local time of MGS-TES observations was fixed to ~14:00 h (daytime), and the background water-vapor column density defined here does not include the diurnal changes. In each run, the variances of $\alpha_{max}$ became close to equilibrium states with diurnal changes (affected by those of surface temperatures) in the end of 20-sol simulation; therefore, we assumed these conditions to be actual states of the grid-mean amount of water vapor emitted from the RSL. Water vapor does not affect the wind fields, but the released water vapor is passively transported by the background flow. The maximum column densities of the emitted water vapor were 2–10 times smaller than $\Delta\alpha$ (Table 1) and 3–4 orders of magnitude smaller than the background water vapor. This is due to horizontal dissipation caused by wind shear around the RSL. Four-dimensional (4-D) variance of the amount and distribution of emitted water vapor around each emission point is shown in Movies S1–S6, together with the surface temperatures, wind fields, and vertical atmospheric instabilities.

The question arises, "how do wind fields and atmospheric instabilities affect the horizontal and vertical diffusions of the emitted water?" Figure 5 shows the simulated vertical distribution of water vapor, together with the potential temperature, at the local time of the end of daytime emission [around 18:00 h for CAP and VM ($L_s$=320°), and around 17:00 h for VM ($L_s$=130°) and the three SML points] on the first sol of the emission. The maximum volume mixing ratios of the emitted water vapor were ~0.2 ppmv for CAP and VM points and ~2 ppmv for SML points, which are smaller than $\Delta q_{v1}$ in Table 1 by three orders of magnitude. This is because of the vertical convection inside the daytime PBLs [also known as convective planetary boundary layer (CPBL)] prior to horizontal advection caused by the wind fields. The emitted water vapor reached up to ~12 km from the surface, with the mixing ratios of >0.01 ppmv at the CAP, SML1, and SML2 points, due to the small gradients of potential temperatures in the daytime PBLs. In contrast, the thicknesses of daytime PBLs were much smaller at the VM and SML3 points, <2 km at VM and ~4 km at SML3. Nevertheless, a part of the water vapor was lifted to ~10 km (altitude) by the effects of mountain-valley winds that blow upward on slopes during the daytime. The typical vertical advection velocity was ~ 2 km h$^{-1}$ or less, and thus, the emitted water vapor was distributed within the PBL within a few hours (Movies S1–S6). Following vertical advection, the wind shear disperses the water-vapor plume horizontally.

Horizontal dissipation due to wind shear is characterized by a typical diffusion timescale of ~1 sol for ~5 × 10$^5$ km$^2$ (Movies S1–S6), which corresponds to an effective diffusion coefficient of ~10$^3$ km$^2$ h$^{-1}$. Figure 6 shows the horizontal distribution of the water-vapor column densities emitted from the RSL in each 20-sol run. The characteristics of each emission run are seen in the broadening of the emitted water. The properties of the maximum grid-mean column density ($\alpha_{max}$) and the sizes of the regions having column densities of >0.5 $\alpha_{max}$ and >0.1 $\alpha_{max}$ ($A_{0.5}$ and $A_{0.1}$, respectively) are summarized in Table 2. $A_{0.1}/A$ ($A$ is the grid size of the emission point, as defined in Equation 1) is the largest for SML3, which shows large differences from the others, because it is located on the east slope of the Hellas basin (the emitted water vapor tends to accumulate in the basin) (Figures 6f and 4f). $A_{0.5}/A$ is the largest for VM ($L_s$=130–140°), and the emitted water from the VM point tends to be accumulated in the valleys (Figure 6b). The characteristics of the advection on VM are quite different in different seasons, seen by the strong southward advection out of the valley ($L_s$=320–331°; Figure 6c). The difference is mainly due to the seasonal change of the wind field in VM, which is dominantly eastward in the former and southward in the latter (see Movies S2 and S3 with wind vectors).

$A_{0.1}/A$ was the smallest for SML1, indicating that most of the emitted water vapor stayed around the emission point (Figure 6d). Movie S4 indicates that some water vapor emitted from SML1 is transported far away from the region of the Olympus mountain in the northern hemisphere. The emission might possibly affect the water conditions in the region, but the amount in our simulation is very small for us to consider this to be plausible. For CAP, the emitted water vapor was first transported westward and then, northward; further, it extended to the northern lowlands over a wide range (Figure 6a, Movie S1). For SML2, the emitted water vapor was transported westward and accumulated in the Argyre Basin, which is ~2000 km away from the emission point (Figure 6e, Movie S5).

Figure 7 shows the ratios of the column densities of the emitted water against the observed background water-vapor column densities obtained using MGS-TES in MY26 (Smith 2008). In the CAP and VM simulations (Figures 7a, 7b, and 7c), $r_{max}$ was ~$3 \times 10^{-4}$. In the SML simulations (Figures 7d, 7e, and 7f), $r_{max}$ exceeded $10^{-3}$ around the emission points, and the maximum value was ~$4.4 \times 10^{-3}$ for SML1. The $r_{max}$ displayed in this study are "smoothed" values within a grid point of the MGCM (>3000 km$^2$); the ratios might be even higher in small craters where the emitted water vapor tends to accumulate (see Section 4).

Finally, we discuss the possibility of condensation of the emitted water vapor. Figure 8 shows the vertical profiles of water-vapor saturation level calculated from the temperature profiles of all local times in our MGCM at each emission point in comparison with the water-vapor mixing ratios taken from the Mars Climate Database (Forget et al.

1999; Navarro et al. 2014) at the corresponding pairs of place and season with the MY26 dust scenario. As seen in the figure, the estimated water-vapor mixing ratio never exceeded the saturation level in the convective PBLs at all emission points. Thus, we can deduce that there is no condensation of the emitted water vapor in PBLs. However, as indicated in Figures 8e and 8f, the surface temperature may be quite cold during nighttime. The edges of the craters possibly create permanent shade, and if such shade exists near RSL, it is possible that the water emitted from these RSL condenses and accumulates as surface ice.

## 4.    Implications for atmospheric observations

The detection limit or uncertainty of the water-vapor column density measurements using the existing remote-sensing observations is a few pr.μm (Smith 2008; Toigo et al. 2013). In contrast, our MGCM simulations showed that, even if the RSL had a wet origin, the accumulation of the released water vapor was limited to a few tens of precipitable nanometers (Table 2). Thus, the released water vapor could not be distinguished from the background level in the scale of the spatial resolution of our GCM simulations (~3000 km$^2$, Table 1). Because of the limited water-vapor accumulation, cloud formation was not expected in our simulations.

The difficulty to detect water vapor released from RSL with atmospheric observations is a robust conclusion even considering the variation in the evaporation rate. We fixed the evaporation rates to be 1 mm h$^{-1}$ and 0.1 mm h$^{-1}$ for pure water and brine, respectively. The evaporation rate of brine varies depending on the salt type and the salinity by a factor of ~5 (Altheide et al. 2009; Chevrier et al. 2009; Hanley et al. 2012).

The temperature dependence of the evaporation rate of pure water leads to the change in the evaporation rate by a factor of ~10 in maximum (Murphy and Koop 2005, here we assumed the temperature variation from 273 K to 310 K, the latter of which is the peak temperature for SML-RSL in our simulations). Because water vapor in our MGCM simulations is transported passively by the background flow, our results are scalable for the change in the evaporation rate; if we increase the evaporation rate by a factor of 10, the resulting water vapor mixing ratio increases by a factor of 10 as well. As we found that the mixing ratio of water vapor released from RSL remains much smaller than that of the background by 3−4 orders of magnitude, the variation in the evaporation rate up to a factor of 10 does not affect our conclusions.

Notably, even without the emission from RSL, the water vapor column density can vary due to the regolith-atmosphere exchange by a few % in a diurnal cycle (Zent et al. 1993; Steele et al. 2017; Savijärvi et al. 2016; 2019; Hu 2019). Thus, this effect further limits the detectability of water vapor evaporation from RSL.

Water-vapor transport in our GCM simulations can provide several insights into the behavior of water vapor transport at a smaller subgrid scale. Figure 9 shows the typical evolution of $\alpha_{max}$, $A_{0.5}$, and $A_{0.1}$ at two different RSL sites (SML2 and VM at $L_s$=320–331°, see Figures S1–S6 for the results of other sites). We found that, in typical cases (Figure 9a), $\alpha_{max}$ reaches its peak value a few hours after the beginning of emission and remains nearly constant during the successive emission, whereas $A_{0.5}$ and $A_{0.1}$ gradually increase. This is caused by rapid vertical advection to the top of the PBL in a few hours and the subsequent horizontal dissipation due to wind shear (Section 3). The cessation of emission leads to a rapid decline in $\alpha_{max}$.

This behavior can be qualitatively modeled using 2D horizontal diffusion with a point source function. During the water-vapor emission, the column density σ is given by:

$$\sigma \sim E \times \Delta t \times A_{RSL}/A_{plume}, \qquad (4)$$

where E, $\Delta t$, $A_{RSL}$, and $A_{plume}$ are the emission rate, time since the emission started, surface area of RSL, and surface area of the extended water-vapor plume (approximated by either $A_{0.5}$ or $A_{0.1}$). Using dimensional analysis, we obtained the value of $A_{plume}$:

$$A_{plume} \sim D \times \Delta t, \qquad (5)$$

where $D$ is the effective diffusion coefficient that determines the horizontal dissipation due to wind shear. Therefore, we deduce the following equation:

$$\sigma \sim E \times \Delta t \times A_{RSL}/(D \times \Delta t) \sim E \times A_{RSL}/D, \qquad (6)$$

which is independent of $\Delta t$.

The above analysis indicates that, as long as the 2D diffusion view is correct, the maximum column density of water vapor in the subgrid scale will reach a steady-state value and will not increase with time; therefore, the water vapor released from RSL in typical cases is not detectable even with high-resolution (RSL scale) nadir observations regardless of the time of day of observations. However, we also note that understanding the transport of water vapor released from RSL on a smaller scale ultimately requires mesoscale modeling (Rafkin et al. 2016; Steele et al. 2017) designed particularly for this purpose. Our simulations partially focused on (a) the wet-origin scenario (Section 2). Notably, because this scenario maximizes the amount of released water vapor, the detectability with atmospheric observations would be lower for (b) the hybrid origin scenario.

Several studies have discussed the possible exchange of atmospheric and subsurface water in RSL sites and other locations. Considering atmospheric water vapor as a possible source of liquid water in RSL, McEwen et al. (2014) compiled the CRISM water-vapor data over a few tens of confirmed RSL. They found no clear correlation with the activity observed in the HiRISE images. Our results suggest that, because of the rapid horizontal dissipation of possible water vapor excess/deficit, their findings do not necessarily rule out the wet origin of RSL. Fouchet et al. (2007) reported an anti-correlation between the normalized water column and the surface pressure for the low- to mid-latitudes and discussed the possible contribution of released subsurface water. Additionally, Leung et al. (2016) proposed that the water vapor released from the RSL may aid the formation of fog in the VM. The sources of water vapor suggested by these studies are possibly related to RSL activities, but our results show that the amount of water required to cause cloud formation is likely far greater than that required to trigger RSL.

Atmospheric transport and dissipation of methane and dust plume released from point sources can be analogs of water vapor emitted from RSL sites. Using mesoscale simulations for dust storms, Spiga et al. (2013) found that the dust disturbance, transported vertically by radiative heating during the daytime, is almost entirely dissipated during the night under the influence of horizontal wind shear; this is consistent with the behavior of water vapor found in our simulations at a larger scale.

Transport of water vapor released from RSL in our simulations is also in general agreement with that of methane, whose transport after the release was simulated in several GCM studies (Lefèvre and Forget 2009; Mischna et al. 2011; Holmes et al. 2015;

2017; Viscardy et al. 2016; Temel et al. 2019). Such behavior includes rapid vertical mixing in PBL, horizontal dissipation to several tens of degrees over a few sols, and the dominance of longitudinal (westward) transport (rather than latitudinal transport). A methane release simulation assuming steady input flux showed the maximum column density reaching a steady-state value by five sols (Mischna et al. 2011), which is consistent with our simulations for steady water vapor release. In contrast to Mischna et al. (2011), Holmes et al. (2017) found a gradual increase in the methane mixing ratio until 10 sols in their sustained emission scenario. The difference between the two studies may be caused by the locality of the methane emission points assumed.

Although our results showed that the water vapor released from RSL rapidly dissipates and does not accumulate in typical cases, we also found that the picture of rapid horizontal diffusion may not be applicable when topography plays an important role. In contrast to SML, where the released water vapor rapidly dissipates during the night (Figure 9a), the results for VM showed moderate accumulation that sustained the night (Figure 9b). The daytime evolution of $\alpha_{max}$, $A_{0.5}$, and $A_{0.1}$ is also inconsistent with simple 2D diffusion. This is caused by topography combined with the low PBL, which limits vertical mixing of water vapor; this has been suggested by a previous study with respect to the Gale crater (Fonseca et al. 2018). Although the topography effect on is insufficient to produce detectable local accumulation of water vapor (with respect to our MGCM's scale), the tendency of RSL to appear preferentially on crater walls indicates that the released water vapor may accumulate locally on a smaller scale. Thus, we suggest that future atmospheric observations dedicated to quantifying water vapor release from RSL should focus on topographic basins (e.g., small craters) that show RSL activities and depths deeper than the PBL during active RSL seasons. Finally, we note that this implication is appreciable not only for RSL but also for any surface water sources, such as exposed ground ice (Dundas et al. 2018). Global and/or mesoscale climate simulations are helpful for identifying such locations.

## 5. Conclusions

In this study, we examined whether atmospheric observations of water vapor and/or water ice clouds can constrain the formation mechanisms of RSL on Mars. We summarized the possible rate and duration of water-vapor emission from RSL in different scenarios, including wet, hybrid, and dry origins. MGCM simulations were performed to observe the water vapor released from several RSL sites (CAP, VM, and SML) transported in the atmosphere. We found that, even in the wet-origin scenario that maximizes the water-vapor emission, the accumulated water vapor was not enough to be distinguished from the background level, with respect to existing observations in typical cases. This is caused by the rapid vertical mixing within the PBL, followed by horizontal dissipation due to wind shear. As a consequence, the water-vapor column densities above the emission points remain nearly constant during the daytime and rapidly decline during the night, which is consistent with the 2D horizontal diffusion. These results suggest that the lack of correlation between the RSL activities and the overlying water-vapor column density does not rule out the possibility of the wet origin of the RSL. Additionally, we found that the proposed signatures of atmosphere-subsurface water exchange, including excess water vapor in the low to mid-latitudes and fog in VM, requires a greater amount of water than that required to trigger RSL activities. We also found that water vapor tends to accumulate in basins/valleys in some cases due to the combined effects of topography and low PBL. Such locations showing RSL activities may be possible targets for future atmospheric observations dedicated to quantifying water-vapor release from RSL to elucidate their formation mechanisms.

**Acknowledgments**

We thank two anonymous reviewers for constructive comments. This study was supported by the Japan Society for the Promotion of Science KAKENHI (Grant numbers 17H06457, 19H01960, 19H05072, 18H04453, 19H00707, 19K03980, 20H04605, 20KK0080, and 21K13976). S. A. is "Chargé de Recherches" of the F.R.S.-FNRS. We thank Dr. M. D. Smith for providing the MGS-TES observational data for surface temperature and water-vapor column density.

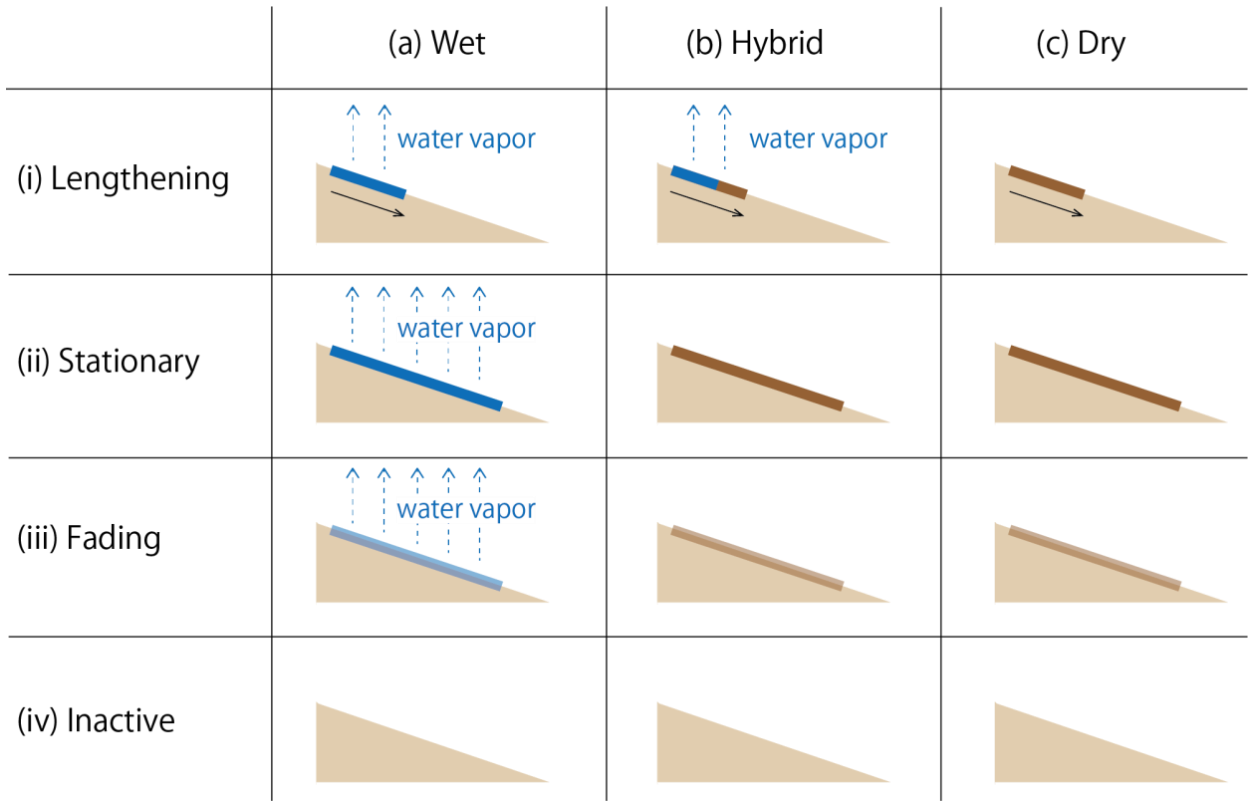

Figure 1: Expected duration (i–iv) of water-vapor release from recurring slope lineae (RSL) for the (a) wet, (b) hybrid, and (c) dry origin scenarios (see text). Blue dashed lines indicate water vapor. Blue and brown areas denote wet and dry RSL, respectively.

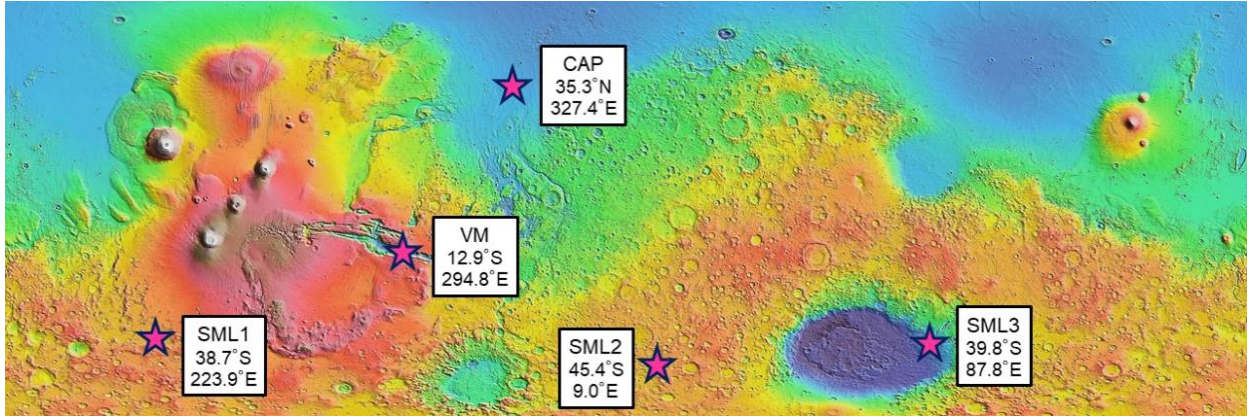

Figure 2: Map of water vapor emission points implemented in the MGCM chosen from observed recurring slope lineae (RSL) (Stillman et al. 2014, 2016, 2017). The background image is the color-coded elevation observed by Mars Orbiter Laser Altimeter (MOLA) onboard MGS (Smith et al., 1999).

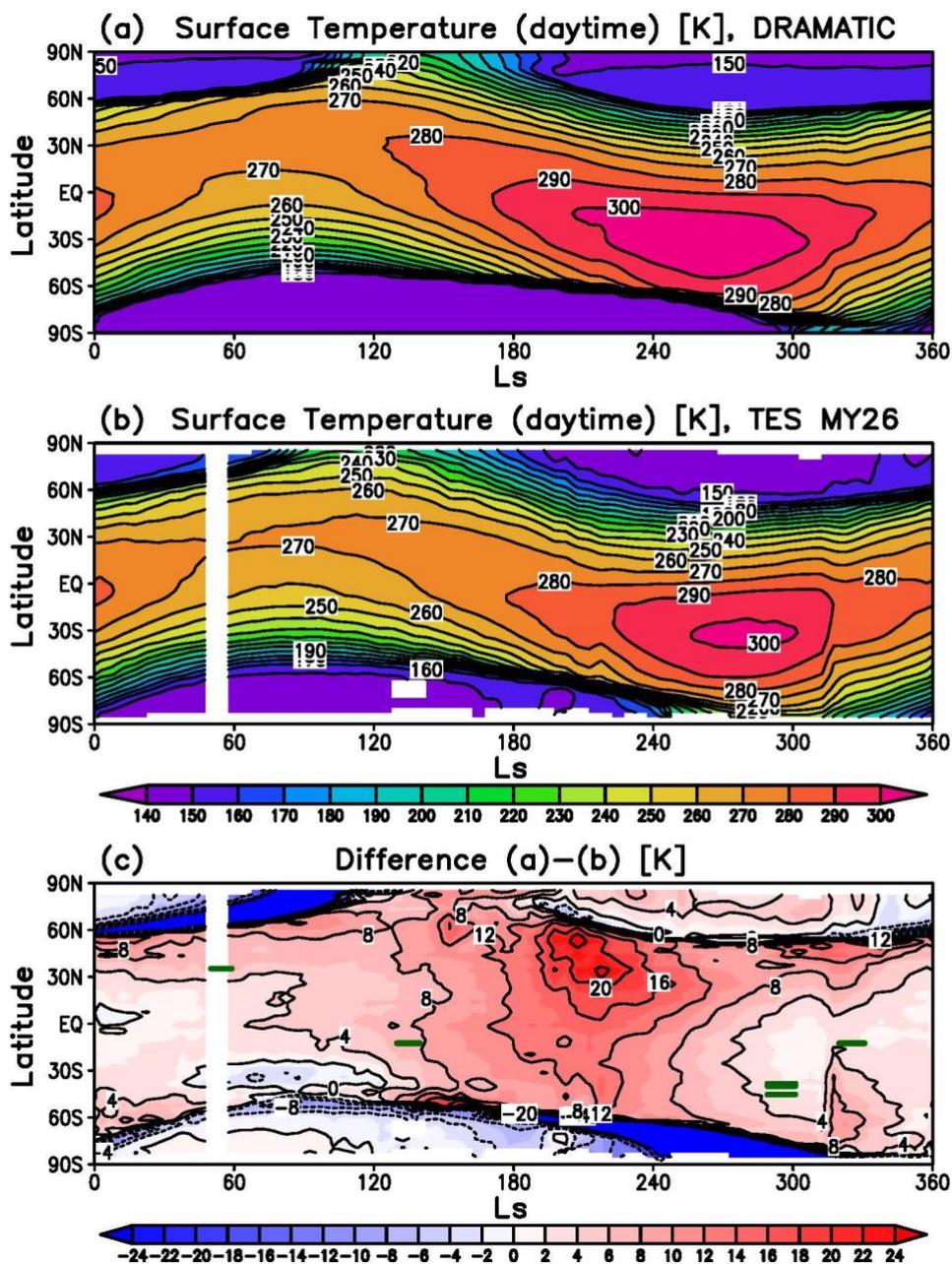

Figure 3: (a) Simulated daytime (~14:00 h) zonal-mean surface temperature in the results of our Mars Global Climate Model (MGCM) with MY26 dust scenario; (b) the same map for Mars Global Surveyor-Thermal Emission Spectrometer (MGS-TES) observations in MY26 (Christensen et al. 2001); (c) differences between the MGCM and observation representing (a) subtracted by (b). Green marks represent the seasons and latitudes corresponding to the water vapor emission points.

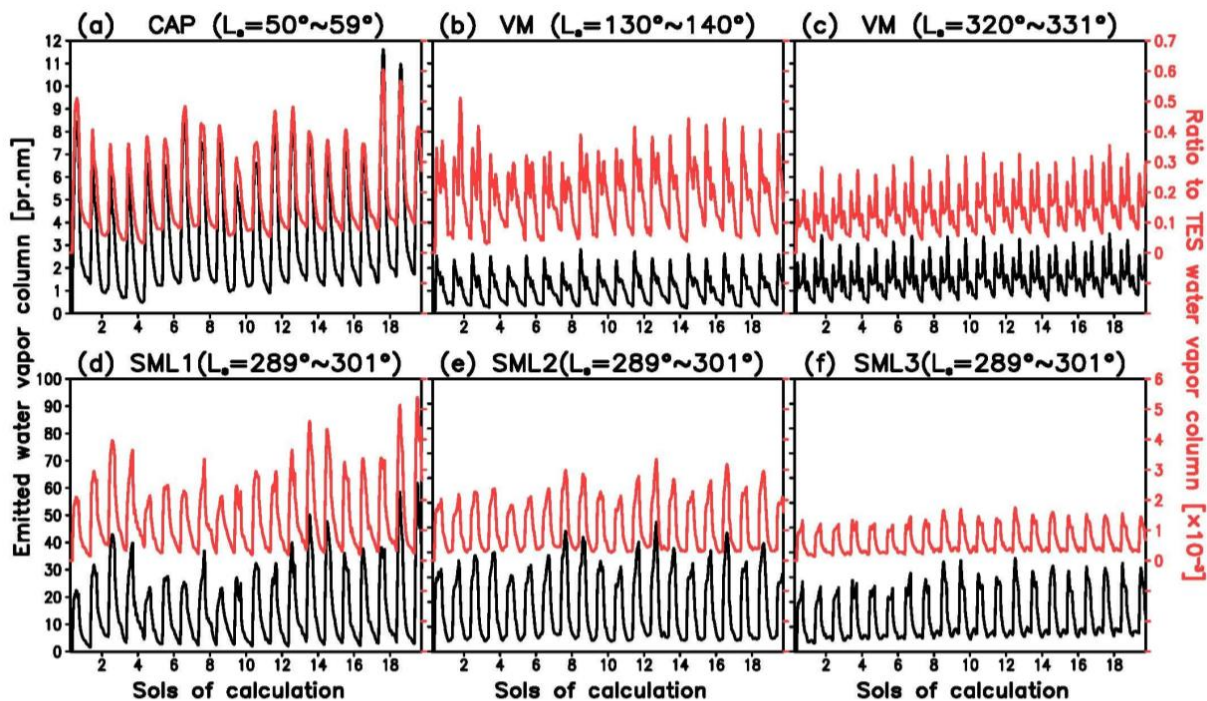

Figure 4: Mars Global Climate Model (MGCM) simulation results. Black lines: maximum grid-mean column densities of the emitted water vapor from RSL (pr.nm), Red lines: their ratios to the observed water-vapor column densities by Mars Global Surveyor-Thermal Emission Spectrometer (MGS-TES) (Smith 2008) at the corresponding season and place, for the six emission runs. (a) CAP ($L_s$=50–59°), (b) VM ($L_s$=130–140°), (c) VM ($L_s$=320–331°), (d) SML1 ($L_s$=289–301°), (e) SML2 ($L_s$=289–301°), and (f) SML3 ($L_s$=289–301°).

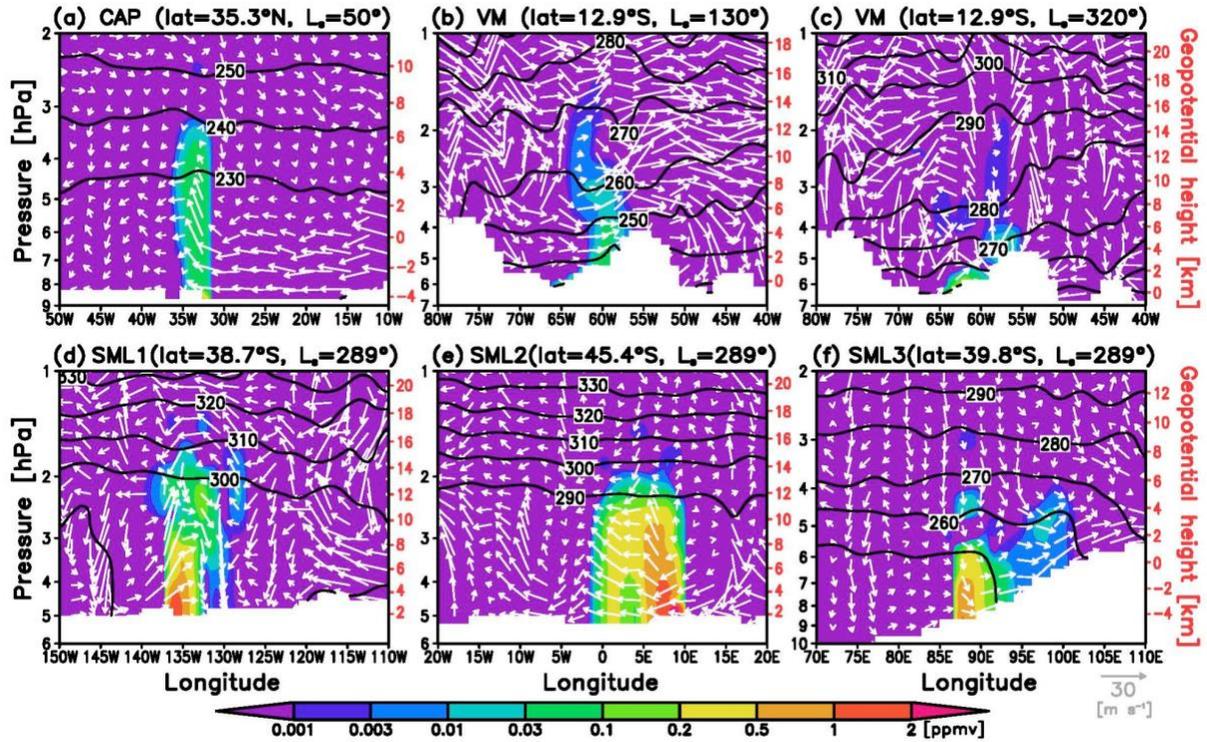

Figure 5: Longitude-altitude cross-sections of (shades) volume mixing ratios of emitted water from RSL, (contours) potential temperatures defined at 6.1 hPa, and (arrows) zonal and vertical wind fields from the Mars Global Climate Model (MGCM) simulations at the latitude of each emission. (a) CAP ($L_s$=50°), (b) VM ($L_s$=130°), (c) VM ($L_s$=320°), (d) SML1 ($L_s$=289°), (e) SML2 ($L_s$=289°), and (f) SML3 ($L_s$=289°). Snapshots at the local time of the end of daytime emission, which is around 18:00 h for (a) and (c), and around 17:00 h for (b), (d), (e), and (f). The results are for the first sol of the emission. The index of zonal wind speed is displayed in the right bottom of the figure, and the vertical wind velocity (shown in the lengths of arrows) is multiplied by 50.

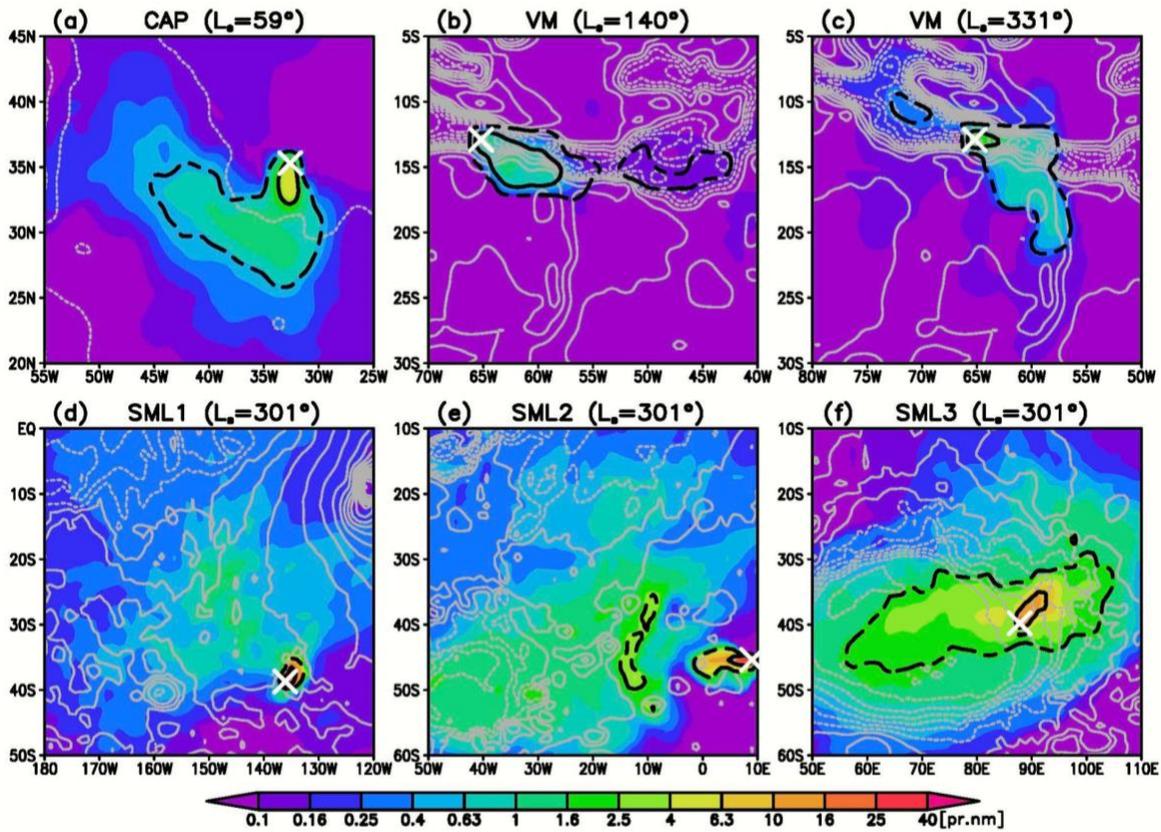

Figure 6: Water vapor column density in our Mars Global Climate Model (MGCM) simulations to show advection of emitted water vapor from RSL for each emission run. (a) CAP ($L_s$=50–59°), (b) VM ($L_s$=130–140°), (c) VM ($L_s$=320–331°), (d) SML1 ($L_s$=289–301°), (e) SML2 ($L_s$=289–301°), and (f) SML3 ($L_s$=289–301°). Shades denote longitude-latitude cross-sections of column densities of emitted water at the local time of the end of daytime emissions on the 20[th] sol of the emission, which is around 18:00 h for (a) and (c), and around 17:00 h for (b), (d), (e), and (f). Black solid and dashed contours denote the regions of the column densities >0.5 and >0.1 times the maximum values, respectively. Gray contours denote the topography (altitude interval of 1 km). White crosses represent the emission points.

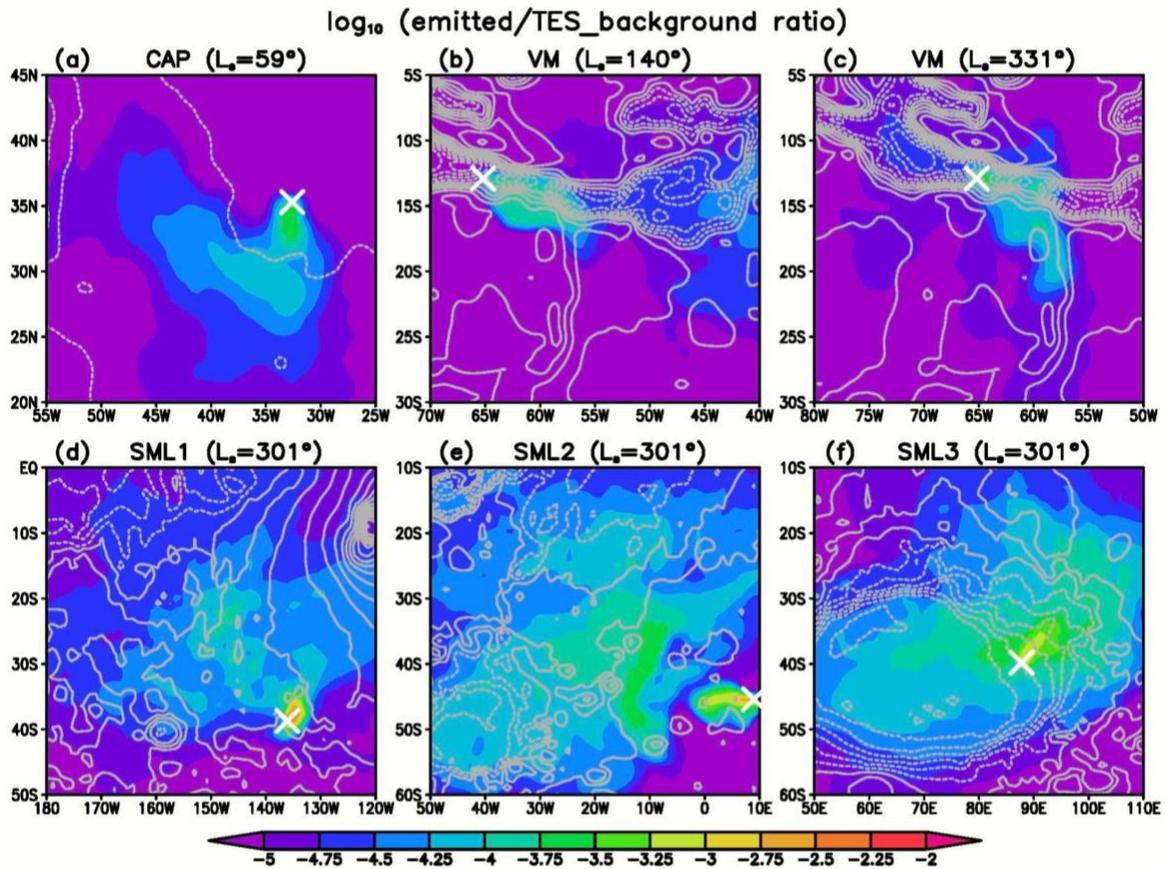

Figure 7: Same as Figure 6, but log₁₀ of their ratios to the background water-vapor column densities taken from the Mars Global Surveyor-Thermal Emission Spectrometer (MGS-TES) observations for corresponding seasons of MY26 (Smith 2008).

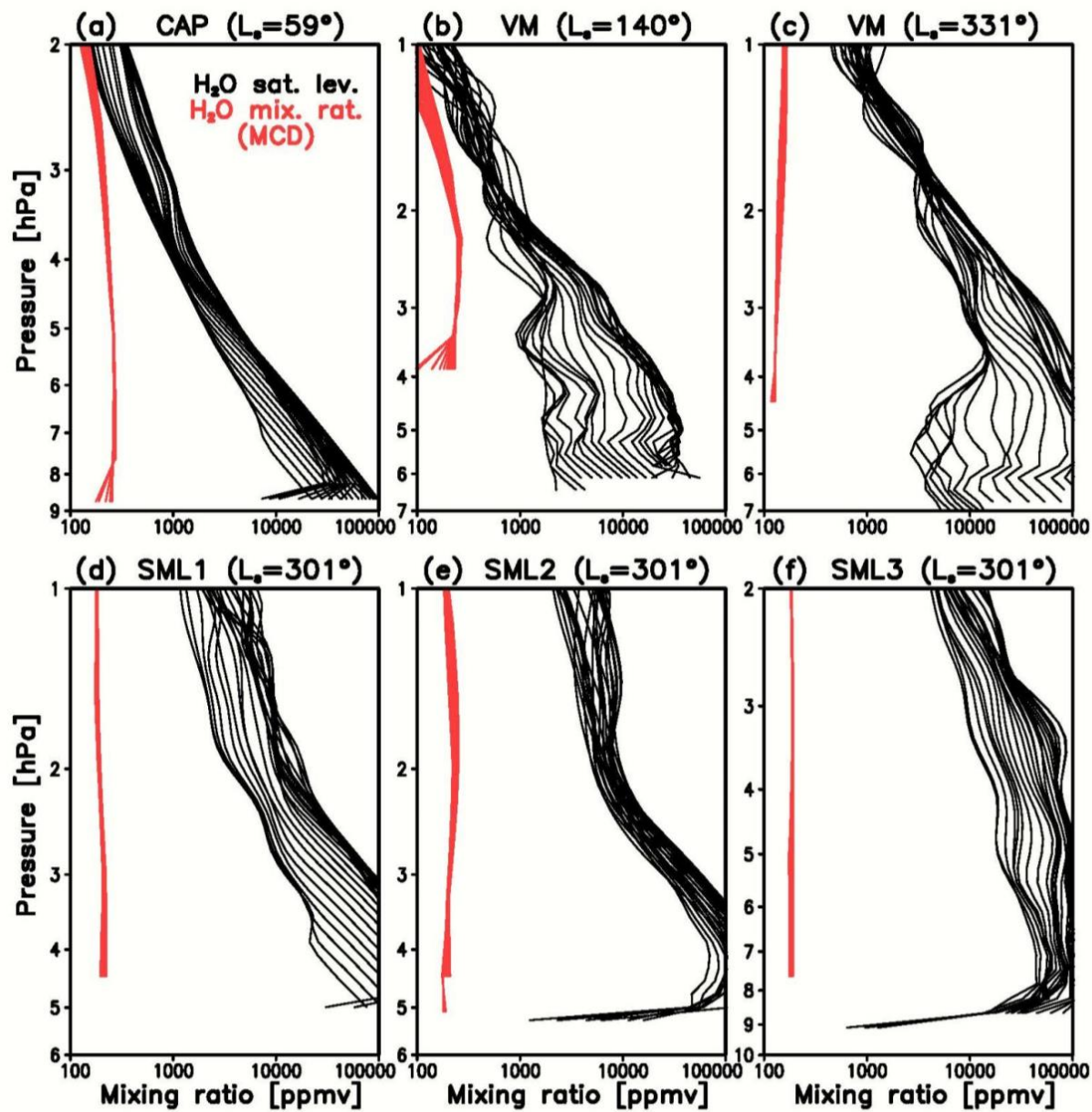

Figure 8: Black lines: Vertical profiles of water-vapor saturation level derived from the vertical temperatures at each emission point in the last 1 sol of each 20-sol emission run in the Mars Global Climate Model (MGCM) simulations (for all local time with 1/2-h interval, 48 profiles); Red lines: vertical profiles of water-vapor mixing ratio in the corresponding place and season (the "background" water vapor) taken from Mars Climate Database (Forget et al. 1999; Navarro et al. 2014) for local time with 1-h interval (24 profiles).

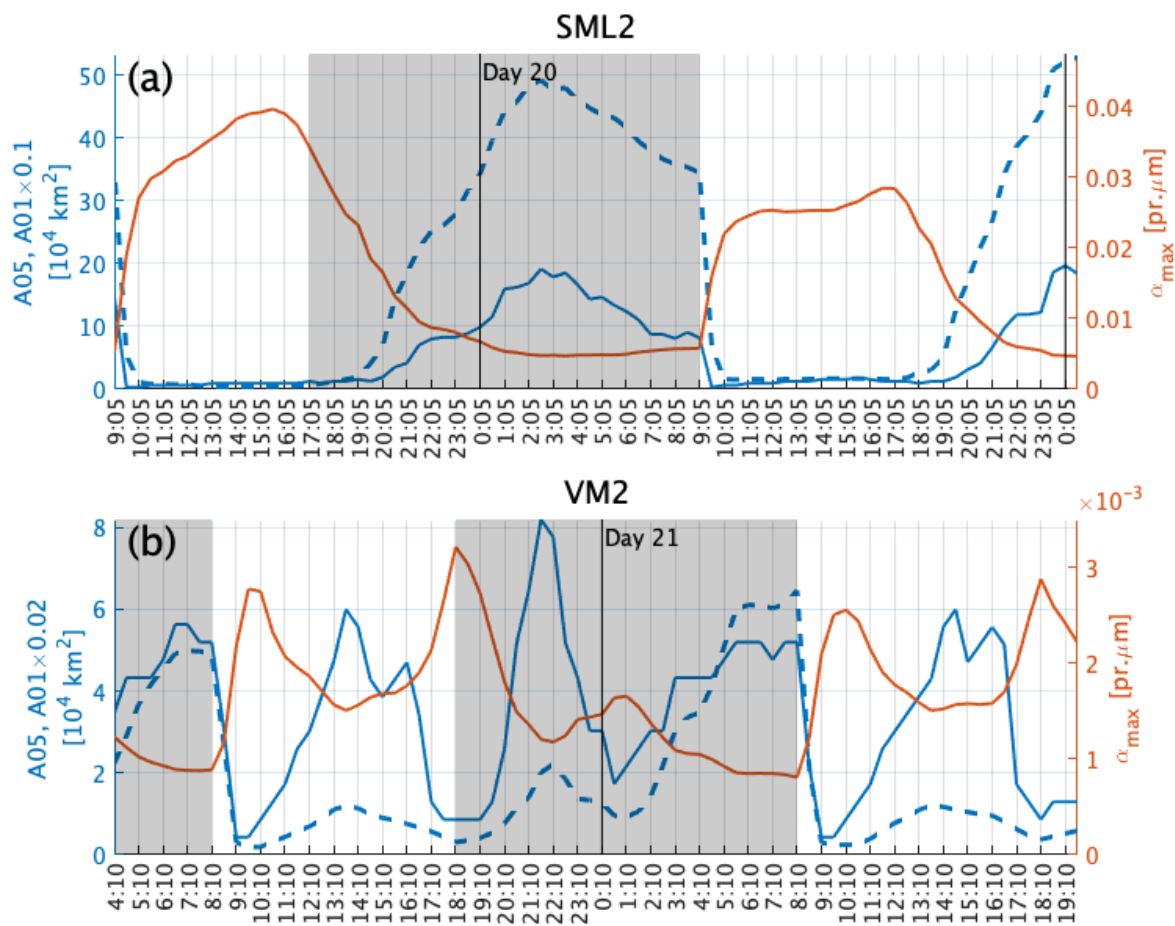

Figure 9: $\alpha_{max}$ (right axis), $A_{0.5}$, and $A_{0.1}$ (left axis, solid, and dashed lines) as a function of time. $A_{0.1}$ values are scaled (×0.1 and ×0.02 for upper and lower panels, respectively). The duration of water-vapor emission is indicated by white areas. The Mars Global Climate Model (MGCM) results for (a) SML2 and (b) VM at $L_s$=320–331° are selected from all periods and places (Figures S1–S6) to show two endmember cases.

| Points and periods | $F_w$ [mm h$^{-1}$] | $A$ [km$^2$] | $p_s$ [Pa] (during $T > T^*$) | $\Delta t$ [h] (per a Sol) | $\Delta \alpha$ [pr.$\mu$m] (per a Sol) | $\Delta q_{v1}$ [ppmv] (per a Sol) |
|---|---|---|---|---|---|---|
| (a) CAP ($L_s$=50°~59°) | 0.1 | 3616 | 880 | 9.88 | 0.027 | 280 |
| (b) VM ($L_s$=130°~140°) | 0.1 | 4321 | 625 | 7.05 | 0.016 | 232 |
| (c) VM ($L_s$=320°~331°) | 0.1 | 4321 | 689 | 9.85 | 0.023 | 299 |
| (d) SML1 ($L_s$=289°~301°) | 1 | 3460 | 513 | 8.50 | 0.246 | 4348 |
| (e) SML2 ($L_s$=289°~301°) | 1 | 3111 | 522 | 8.18 | 0.263 | 4569 |
| (f) SML3 ($L_s$=289°~301°) | 1 | 3405 | 905 | 7.68 | 0.226 | 2265 |

Table 1: Conditions of the Mars Global Climate Model (MGCM) simulations of water-vapor emission from RSL and estimated $\Delta \alpha$ and $\Delta q_{v1}$ from Equations 2 and 3 for the six simulations. $p_s$ and $\Delta t$ are the averaged values in the 20-sol runs. The displayed $A$ values are those obtained using the MGCM, not the derived values from Equation 1.

| Points and periods | $\alpha_{max}$ [pr.nm] | $r_{max}$ | $A_{0.5}$ [km$^2$] | $A_{0.5}/A$ | $A_{0.1}$ [km$^2$] | $A_{0.1}/A$ |
|---|---|---|---|---|---|---|
| (a) CAP ($L_s$=50°~59°) | 5.55 | $3.138\times10^{-4}$ | $1.100\times10^4$ | 3.04 | $2.691\times10^5$ | 74.42 |
| (b) VM ($L_s$=130°~140°) | 1.21 | $1.391\times10^{-4}$ | $7.716\times10^4$ | 17.85 | $2.352\times10^5$ | 54.44 |
| (c) VM ($L_s$=320°~331°) | 2.88 | $1.562\times10^{-4}$ | $8.641\times10^3$ | 2.00 | $1.879\times10^5$ | 43.49 |
| (d) SML1 ($L_s$=289°~301°) | 50.52 | $4.399\times10^{-3}$ | $1.043\times10^4$ | 3.02 | $3.507\times10^4$ | 10.14 |
| (e) SML2 ($L_s$=289°~301°) | 28.44 | $2.092\times10^{-3}$ | $1.244\times10^4$ | 4.00 | $1.576\times10^5$ | 50.67 |
| (f) SML3 ($L_s$=289°~301°) | 17.69 | $8.354\times10^{-4}$ | $4.917\times10^4$ | 14.44 | $1.381\times10^6$ | 405.6 |

Table 2: Horizontal extent of emitted water vapor in the Mars Global Climate Model (MGCM) simulations. The values are based on the snapshots at the local time of the end of daytime emissions, which is around 18:00 h for (a) and (c), and around 17:00 h for (b), (d), (e), and (f) on the 20[th] sol of the emission runs. $\alpha_{max}$ is the maximum grid-mean column density of emitted water in each snapshot, $r_{max}$ is the ratio of $\alpha_{max}$ to the water-vapor column density observed by Mars Global Surveyor-Thermal Emission Spectrometer (MGS-TES) for corresponding place and season in MY26 (Smith 2008), and $A_{0.5}$ and $A_{0.1}$ are the sizes of the regions with column densities of >0.5$\alpha_{max}$ and >0.1$\alpha_{max}$, respectively. $A$ is the grid size of each emission point as defined in Table 1.

# Supplementary Materials

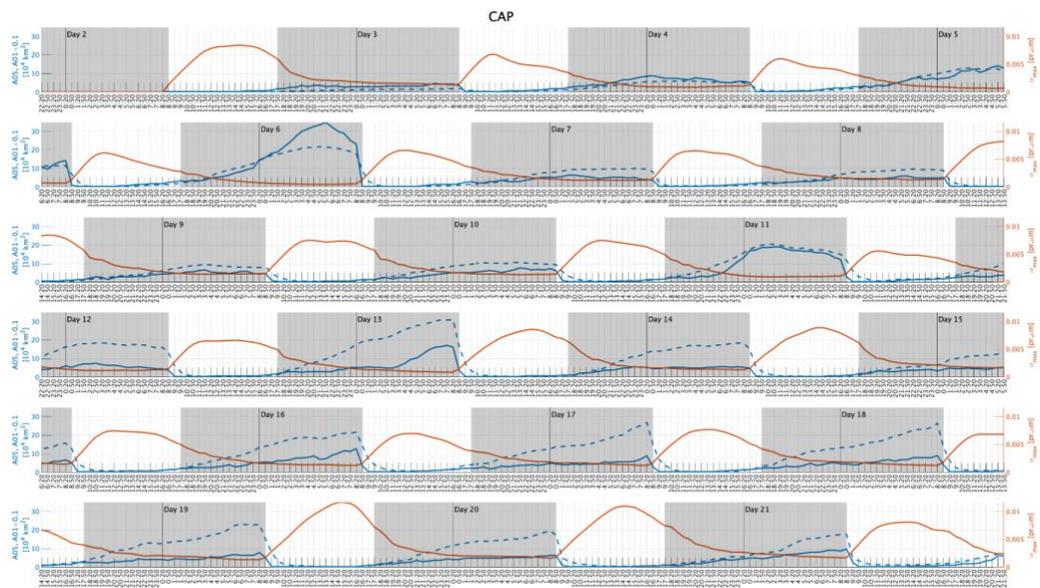

Figure S1: $\alpha_{max}$ (right axis), $A_{0.5}$, and $A_{0.1}$ (left axis) as a function of time (CAP).

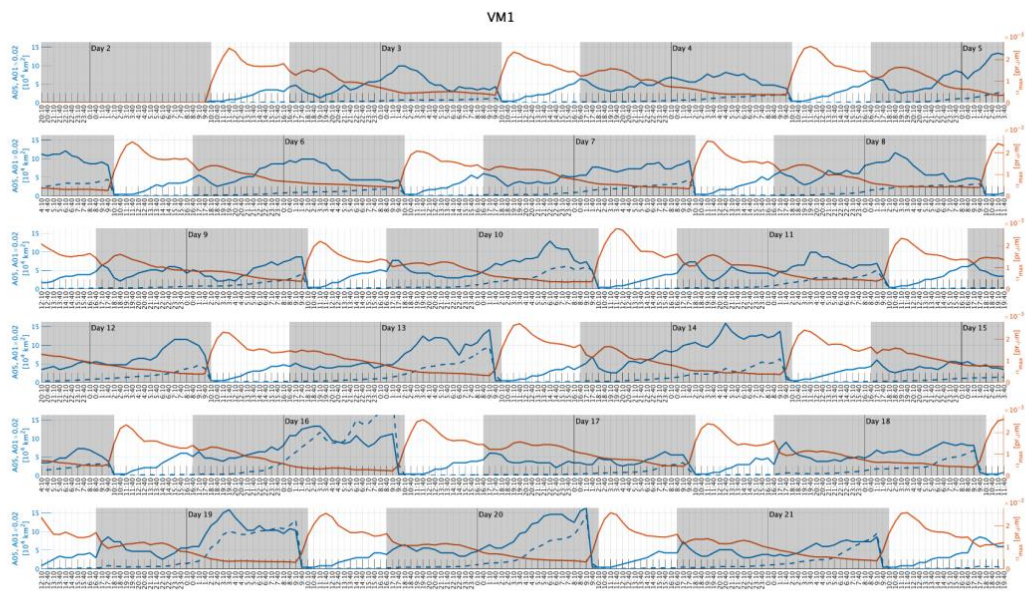

Figure S2: $\alpha_{\max}$ (right axis), $A_{0.5}$, and $A_{0.1}$ (left axis) as a function of time (VM, $L_s$=130°–140°).

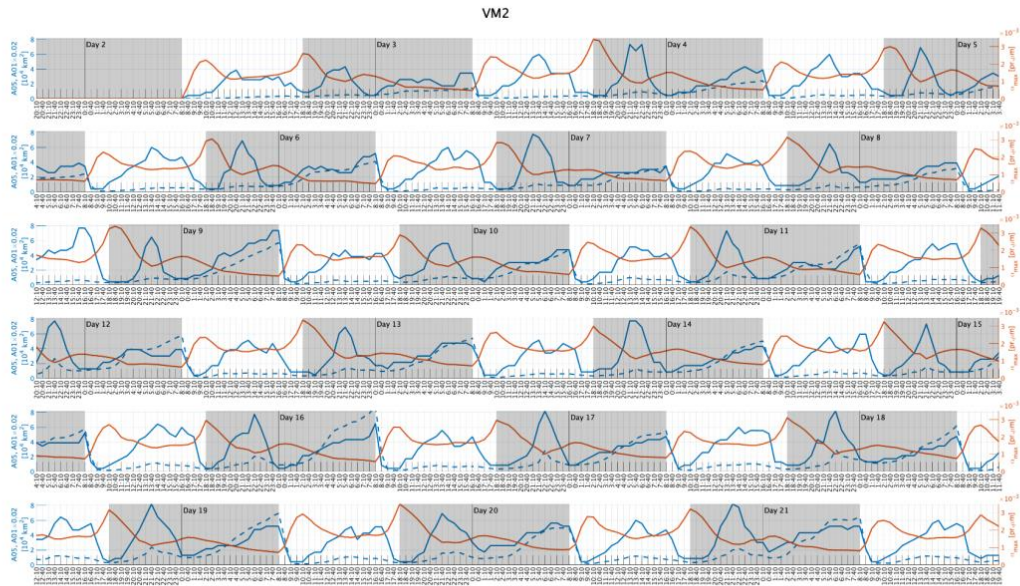

Figure S3: $\alpha_{\mathrm{max}}$ (right axis), $A_{0.5}$, and $A_{0.1}$ (left axis) as a function of time (VM, $L_{\mathrm{s}}$=320°–331°).

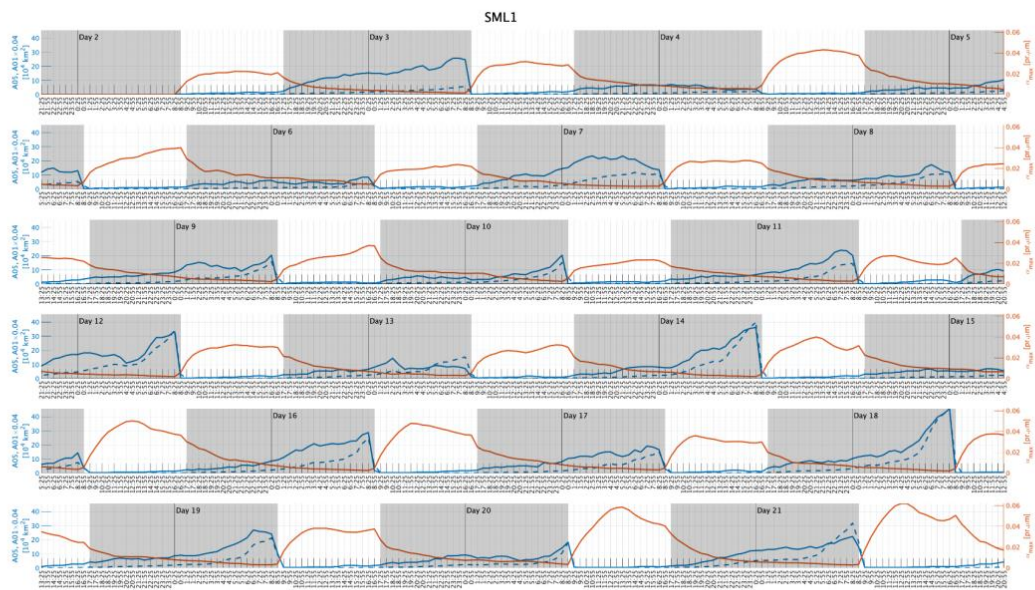

Figure S4: $\alpha_{\max}$ (right axis), $A_{0.5}$, and $A_{0.1}$ (left axis) as a function of time (SML1).

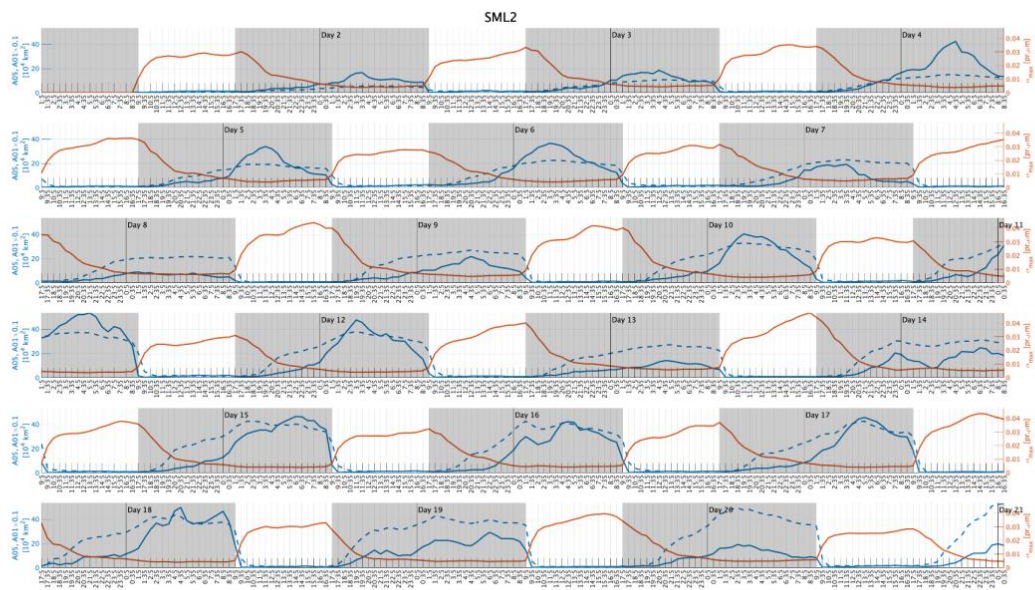

Figure S5: $\alpha_{max}$ (right axis), $A_{0.5}$, and $A_{0.1}$ (left axis) as a function of time (SML2).

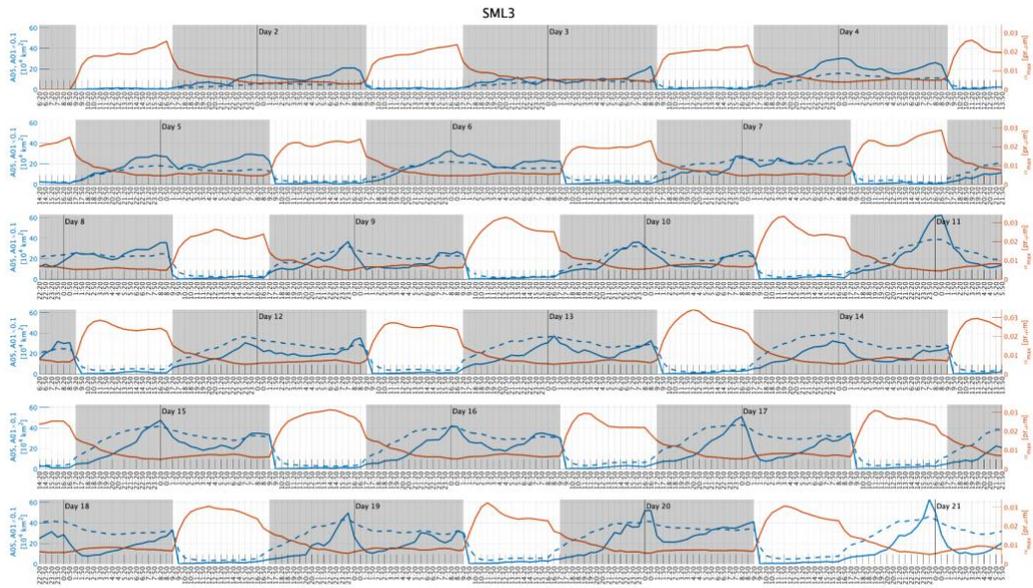

Figure S6: $\alpha_{max}$ (right axis), $A_{0.5}$, and $A_{0.1}$ (left axis) as a function of time (SML3).

Movie S1: Distribution of water vapor emitted from the CAP emission point and temperature/wind fields around the point starting from $L_s = 50°$ for 20 sols (top left). Color shades denote the column density of the emitted water (pr.nm), and arrows denote the wind field at an altitude of ~5 m. The crossover point of the two white lines represents the emission point, and the black contours denote the topography (top right). Color shades denote the surface temperature, and the thick black contour represents the threshold temperature of emission (238 K). The gray contours denote the topography (bottom row), Longitude-altitude (left), and latitude-altitude (right) cross-sections of the mixing ratio of emitted water (ppbv) (color shades), potential temperature defined at 6.1 hPa (black contours), and horizontal (zonal or meridional) -vertical wind fields (arrows). The index of zonal wind speed is displayed at the bottom right, and the vertical wind velocity (shown in the lengths of the arrows) is multiplied by 50.

Movie S2: Same as Movie S1, except that emission from the VM point started from $L_s$=130°.

Movie S3: Same as Movie S1, except that emission from the VM point started from $L_s$=320°.

Movie S4: Same as Movie S1, except emission from the SML1 point started from $L_s$=289°, and the threshold temperature of emission represented by the thick black contour in the top-right column is 273 K.

Movie S5: Same as Movie S4, except the SML2 emission point.

Movie S6: Same as Movie S4, except the SML3 emission point.